\RequirePackage[orthodox,l2tabu]{nag}

\documentclass[a4paper,11pt]{article}
\usepackage[T1]{fontenc}        
\usepackage[utf8]{inputenc}
\usepackage[english]{babel}

\usepackage[left=2.7cm,
            right=2.7cm,
            top=2.7cm,
            bottom=3.2cm
            ]{geometry}        

\usepackage[protrusion,
            expansion
            ]{microtype} 

\usepackage{booktabs}    

\usepackage[font=small,labelfont=bf]{caption}
\usepackage{subcaption}

\usepackage{enumitem}

\usepackage{authblk}    
\usepackage{appendix}
\usepackage{fancyhdr}

\usepackage{braket}
\usepackage{xspace}
\usepackage{mathdots}
\usepackage{stackrel}
\usepackage{xcolor}
\usepackage{mathtools}
\usepackage{graphicx}
\usepackage{amsmath,          
            amssymb,          
            amsthm}          
\usepackage{xfrac}
\usepackage{comment}

\usepackage{mathrsfs}
\usepackage{bbm}      
\usepackage{tikz}

\usepackage[sort&compress, numbers, square]{natbib}

\definecolor{mycolor1}{HTML}{1F7A8C}
\definecolor{mycolor2}{HTML}{2D3362}
\definecolor{mycolor3}{HTML}{BF1363}
\definecolor{mycolor4}{HTML}{F39237}
\definecolor{mycolor5}{HTML}{2D8B6A}

\usepackage[colorlinks,
            linkcolor=black,
            citecolor=black,
            urlcolor=mycolor3,
            bookmarks,
            bookmarksnumbered
            ]{hyperref}    

\usepackage{cleveref}
\crefname{figure}{Fig.}{Figs.}
\crefname{equation}{Eq.}{Eqs.}
\crefname{section}{Sec.}{Secs.}

\allowdisplaybreaks

\theoremstyle{definition}

\theoremstyle{remark}

\newcommand{\RR}{\mathbb{R}}
\newcommand{\ZZ}{\mathbb{Z}}

\newcommand{\ii}{\mathrm{i}}

\newcommand{\dd}{\mathrm{d}}

\newcommand{\ctm}{\widehat{A}}

\newcommand{\abs}[1]{\left\vert#1\right\vert}
\newcommand{\avg}[1]{\langle #1 \rangle}
\newcommand{\nord}[1]{:\! #1\! :}   

\newcommand{\lr}[1]{\left( {#1} \right)} 
\newcommand{\slr}[1]{\left[{#1} \right]}

\DeclareMathOperator{\Tr}{Tr}   

\title{\textbf{Entanglement Hamiltonian in the  non-Hermitian SSH model}}

\author[1]{Federico Rottoli}
\author[1]{Michele Fossati}
\author[1,2]{Pasquale Calabrese}

\affil[1]{\textit{SISSA and INFN Sezione di Trieste, via Bonomea 265, 34136 Trieste, Italy.}}
\affil[2]{\textit{ICTP, Strada Costiera 11, 34151 Trieste, Italy.}}

\date{}

\begin{document}
\maketitle

\begin{abstract}
Entanglement Hamiltonians provide the most comprehensive characterisation of entanglement in extended quantum systems.
A key result in unitary quantum field theories is the Bisognano-Wichmann theorem, which establishes the locality of the entanglement Hamiltonian.
In this work, our focus is on the non-Hermitian Su–Schrieffer–Heeger (SSH) chain. 
We study the entanglement Hamiltonian both in a gapped phase and at criticality.
In the gapped phase we find that the lattice entanglement Hamiltonian is compatible with a lattice Bisognano-Wichmann result, with an entanglement temperature linear in the lattice index.
At the critical point, we identify a new imaginary chemical potential term absent in unitary models. 
This operator is responsible for the negative entanglement entropy observed in the non-Hermitian SSH chain at criticality.

\end{abstract}

\tableofcontents
\newpage

\section{Introduction}\label{sec:intro}

In recent years, the study of entanglement has attracted a lot of interest from several different communities and has emerged as a unifying theme across quantum physics, in fields ranging from quantum information~\cite{Nielsen:2012yss} and high energy physics~\cite{Nishioka:2009un}, to statistical mechanics~\cite{Amico:2007ag, Calabrese:2009kka, Eisert:2008ur} and condensed matter physics~\cite{Laflorencie:2015eck}.
Given a pure state described by the density matrix $\rho = \ket{\Psi}\bra{\Psi}$ and considering a bipartition of the system in $A$ and $B$, the information about the entanglement between the two subsystems is encoded in the reduced density matrix, obtained by tracing over the Hilbert space of one of the two subsystems
\begin{equation}\label{eq:rhoA}
    \rho_A = \Tr_B \rho.
\end{equation}
If the density matrix $\rho$ is entangled, then the reduced density matrix in \cref{eq:rhoA} corresponds a mixed state and the von Neumann and R\'enyi entropies of $\rho_A$~\cite{Calabrese:2004eu, Calabrese:2009qy}
\begin{equation}\label{eq:ee}
    S_A = - \Tr \slr{ \rho_A \log \rho_A }, \qquad S_A^{(n)} = \frac{1}{1-n} \log \Tr \rho_A^n.
\end{equation}
are good entanglement monotones.
The knowledge of all the R\'enyi entropies allows in turn to compute the full entanglement spectrum, i.e, the spectrum of the reduced density matrix~\cite{Li:2008kda, Calabrese:2008iby, Alba:2017bgn}.

While the entanglement entropies in \cref{eq:ee} are very useful, they do not entirely capture the entanglement properties of the system.
Over the years, more comprehensive characterisations, beyond what can be exclusively derived from the knowledge of the entanglement spectrum, have been investigated.
Arguably, the most complete understanding stems from the entanglement (or modular) Hamiltonian (EH), which is the logarithm of the (normalised) reduced density matrix~\cite{Haag:1996hvx, Witten:2018zxz, Dalmonte:2022rlo}
\begin{equation}\label{eq:eh}
    \rho_A = \frac{e^{-K_A}}{Z_A}.
\end{equation}
This operator contains much more information than the entanglement entropies, since, unlike the latter, its specific form depends not only on the eigenvalues but also on the eigenvectors of $\rho_A$.

Although calculating the EH is considerably more challenging than determining the entanglement entropies, a rich theoretical framework has been developed during the last decades.
The most remarkable result  is the Bisognano-Wichmann (BW) theorem~\cite{Bisognano:1975ih, Bisognano:1976za, Haag:1996hvx, Witten:2018zxz, Dalmonte:2022rlo}.
Considering the vacuum state of a unitary relativistic quantum field theory (QFT) on $\RR^{d+1}$ and taking as subsystem the half-space $x^1>0$,
the Bisognano-Wichmann theorem asserts that the entanglement Hamiltonian in the half-space is the generator of Lorentz boosts
\begin{equation}\label{eq:BW}
    K_A = 2 \pi \int_{x^1 > 0} \dd^d x\, x^1\, T_{00}\!\left(x\right),
\end{equation}
where $T_{00}(x)$ is the energy density.
This result is remarkable for several reasons. 
First, it is extremely general, holding for all unitary Lorentz invariant QFTs, independently of the dimension of space-time and of the mass spectrum.
Another significant property is the fact that the BW Hamiltonian in \cref{eq:BW} has a local structure, given by the integral of local operator with a linearly increasing local entanglement temperature $\beta(x) = x^1$.
Finally, the Bisognano-Wichmann theorem provides a mathematical proof of the Unruh effect~\cite{Fulling:1972md, Davies:1974th, Unruh:1976db}, as discussed in~\cite{Haag:1996hvx, Witten:2018zxz, Dalmonte:2022rlo}.

For generic massive QFTs, the Bisognano-Wichmann theorem in \cref{eq:BW} is the only known analytic result.
For conformal field theories (CFT), instead, the extended symmetry makes it possible to obtain more general results.
In particular, the Hislop-Longo theorem~\cite{Hislop:1981uh, Haag:1996hvx} (see also~\cite{Casini:2011kv, Wong:2013gua, Cardy:2016fqc}) provides the entanglement Hamiltonian of the ground state of a CFT in any ball shaped region.
Even more general results can be obtained in $1+1$-dimensional unitary CFTs, where the infinite dimensional Virasoro symmetry allows one to map the Bisognano-Wichmann result in \cref{eq:BW} in several different geometries~\cite{Wong:2013gua, Cardy:2016fqc}.
In particular, for the conformal vacuum state, if the subsystem is a single interval ($A=[0,\ell]$) the entanglement Hamiltonian takes the form~\cite{Hislop:1981uh, Casini:2011kv, Wong:2013gua, Cardy:2016fqc}
\begin{equation}\label{eq:CFTinterval}
    K_A = 2\pi\int_0^\ell \dd x\, \beta\!\lr{x} T_{00}\!\lr{x}, \qquad \text{with }\, \beta(x) = \frac{x \left ( \ell - x \right )}{\ell}\,,
\end{equation}
a formula that inherits the local structure of the Bisognano-Wichmann result, with a parabolic local entanglement temperature $\beta(x)$.
We remark that this local structure is very peculiar and it fails to hold as soon as we consider minor modifications such as  for the lowest excited states~\cite{Sarosi:2017rsq} and when considering multiple intervals~\cite{Casini:2009vk, Arias:2018tmw}.
This property is not only of theoretical interest, but recently in Refs.~\cite{Dalmonte:2017bzm, Kokail:2020opl, Kokail:2021ayb, Zache:2021yju, Joshi:2023rvd} it has been leveraged to efficiently construct in synthetic quantum systems the ground state of lattice models using a variational approach.

A separate line of research, yielding numerous results, concerns the study of the EH in unitary $1+1$-dimensional integrable lattice model~\cite{pescheltruong1987, Davies:1988zz, truongpeschel1989, peschelkaulkelegeza, Davies:1989zz, Frahm:1991hr, Tetelman, Thacker:1985gz, Ercolessi:2009kc, pescheltruong1991, peschelchung1999, Eisler:2020lyn, Giudici:2018izb, Mendes-Santos:2019tmf, Zhang:2020mjv, Eisler:2023yys}.
In these systems, the entanglement Hamiltonian of the ground state in the half-space $x>0$ is intimately related to Baxter's corner transfer matrix (CTM) $\ctm$~\cite{BaxterBOOK, Baxter:1976uh, Baxter:1977ui}.
Considering, for example, isotropic square lattices, the effect of the corner transfer matrix is to add a full angular segment to a piece of lattice, mapping a horizontal row to a vertical one and vice versa.
Using this property it is possible to show that the lattice reduced density matrix in the half-line can be expressed as the product of four CTMs~\cite{nishino1997corner, peschelkaulkelegeza,Calabrese:2004eu}
\begin{equation}\label{eq:rhoACTM}
    \rho_A = \frac{\ctm^4}{\Tr \ctm^4},
\end{equation}
where $Z = \Tr \ctm^4$ is the partition function.
Recalling the definition \eqref{eq:eh} of the entanglement Hamiltonian, \cref{eq:rhoACTM} implies that it is proportional to the logarithm of fourth power of the corner transfer matrix~\cite{peschelkaulkelegeza}
\begin{equation}
    K_A = - \log \ctm^4.
\end{equation}

This correspondence between entanglement Hamiltonians and CTMs has made it possible to obtain the EHs in several integrable models.
It has been observed that in certain integrable models, the logarithm of the CTM and the EH can be written in terms of the density of the lattice Hamiltonian $h_{j}$ with a linearly increasing local temperature
\begin{equation}\label{eq:TTI}
    K_A \propto \sum_{j = 0}^{\infty} j\, h_{j}\,,
\end{equation}
with a non-trivial proportionality constant.
This behaviour has been identified in various spin systems such as the Ising model~\cite{Davies:1988zz, truongpeschel1989, peschelkaulkelegeza}, the XXZ~\cite{BaxterBOOK, Davies:1989zz, Frahm:1991hr}, the XYZ chains~\cite{Tetelman, Thacker:1985gz, Ercolessi:2009kc}, the anisotropic XX chain~\cite{Eisler:2020lyn}, and in bosonic models such as the harmonic chain~\cite{pescheltruong1991, peschelchung1999, Eisler:2020lyn}.
Comparing \cref{eq:TTI} with the Bisognano-Wichmann theorem in \cref{eq:BW}, it is evident that the two entanglement Hamiltonians share the same structure.
In fact, the connection between the two results runs deeper than a superficial similarity.
Tetel'man~\cite{Tetelman} and Itoyama and Thacker~\cite{Thacker:1985gz, Itoyama:1986ad, Thacker:1988vu, Itoyama:1988zn} independently showed that in these integrable models the logarithm of the corner transfer matrix is the generator of a continuous group of lattice Lorentz transformations, akin to the role played by the generator of Lorentz boosts in the BW theorem.

Despite the wealth of results for unitary models, nothing is known for non-Hermitian theories.
In particular, since one of the hypothesis of the Bisognano-Wichmann theorem \eqref{eq:BW} is that the Hilbert space carries a unitary representation of the Poincar{\'e} group~\cite{Haag:1996hvx, Bisognano:1975ih, Bisognano:1976za}, it is not obvious how to adapt  to non-unitary CFTs this theorem and its corollary  \eqref{eq:CFTinterval}.
Non-Hermitian models~\cite{MoiseyevBOOK, Ashida:2020dkc} have recently attracted a lot of interest for several reasons, including but not restricted to the study of the $PT$-symmetric systems~\cite{Bender:1998ke, Bender:2015aja, El-Ganainy:2018ksn}, optical phenomena~\cite{Feng:2017hot, mirialu2019} and the study of open systems~\cite{graefekorschnederle, Rotter:2009zhs, Muller:2012bxu} and measurement induced transitions~\cite{Gopalakrishnan:2020atr, biellaschiro2021, Turkeshi:2021mcz, Muller:2021xxc, Turkeshi:2022ipq}. 
It is then very natural to explore the entanglement properties within this class of systems.

A pioneering study was carried out in Ref.~\cite{ChangRyu:2019}, where the authors have studied the entanglement entropy and the entanglement spectrum in the non-Hermitian Su-Schrieffer-Heeger (SSH) model at criticality (reviewed in \cref{sec:model}).
Remarkably, it was observed that the entanglement entropies obey the logarithmic dependence on the subsystem length typical of critical systems~\cite{Holzhey:1994we, Calabrese:2004eu}, but with a negative central charge $c=-2$ (see also~\cite{Tu:2021xje}).
Later, in Ref.~\cite{Fossati:2023zyz}, the analysis has been extended to the symmetry resolved EEs.
In this work we move a step further, conducting an exploratory and thorough numerical investigation of the entanglement Hamiltonian in the non-Hermitian Su-Schrieffer-Heeger model, both in the gapped phase and at criticality.
In the gapped phase we observe that the lattice EH has a structure analogous to the one of integrable lattice models reported in \cref{eq:TTI}.
At the critical point, we instead find an additional term not accounted for in the Bisognano-Wichmann corollary in \cref{eq:CFTinterval}, which is responsible for the negativeness of the entanglement entropies.

The present manuscript is organised as follows. First, in \cref{sec:model} we review the non-Hermitian Su-Schrieffer-Heeger model, with particular focus on the non-unitary $c=-2$~$bc$-ghost CFT which describes the critical point.
In \cref{sec:EH} we report the main results of this work, the numerical lattice entanglement Hamiltonian in the non-Hermitian SSH model.
We first consider the topologically trivial gapped phase in \cref{sec:EHgapped} and we then study the critical point in \cref{sec:EHgapless}.
We draw our conclusions in \cref{sec:conclusions}.

\section{The non-Hermitian Su–Schrieffer–Heeger model}\label{sec:model}

\begin{figure}
\centering
\begin{tikzpicture}
    \def \a {1.5cm}
    \def \rad {1.6pt}

{ \footnotesize
    \node[below] at (-3*\a,0) {$0$};
    \node[below] at (-2*\a,0) {$1$};
    \node[below] at (-1*\a,0) {$2$};
    \node[below] at (0*\a,0) {$\cdots$};
    \node[below] at (1*\a,0) {$\ell$};
    \node[below] at (2*\a,0) {$\cdots$};
    \node[below] at (3*\a,0) {$L-1$};
}
\large
    \foreach \j in {-3,...,2}
    {
        \pgfmathtruncatemacro{\iseven}{mod(\j,2)}
        \ifnum\iseven=0
            \draw[mycolor1, very thick] (\j*\a+\rad+0.2,0) -- (({(\j+1)*\a-\rad-0.2},0);
        \else
            \draw[mycolor2, very thick] (\j*\a+\rad+0.2,0) -- ({(\j+1)*\a-\rad-0.2},0);
        \fi
    }

    \foreach \j in {-3,...,3}
    {
        \node[circle, mycolor3, fill, inner sep=\rad] at (\j*\a,0) {};
        \node[circle, fill, inner sep=\rad/2] at (\j*\a,0) {};
    }

    \foreach \j in {-2,...,1}
    {
        \pgfmathtruncatemacro{\iseven}{mod(\j,2)}
        \ifnum\iseven=0
            \ifnum\j>-2
                \node[above, mycolor3] at (\j*\a,0) {$-\ii u$};
            \fi
            \node[above, mycolor1] at ({(\j+0.5)*\a},0) {$v$};
        \else
            \node[above, mycolor3] at (\j*\a,0) {$\ii u$};
            \node[above, mycolor2] at ({(\j+0.5)*\a},0) {$w$};
        \fi
    }
    
\end{tikzpicture}
\caption{Schematic representation of the nH-SSH model, described by \cref{eq:SSHhamiltonian}. The nearest neighbours hoppings have alternating strengths $v$ and $w$. The imaginary chemical potential is set to $\ii u$ on the even sites and $-\ii u$ on the odd sites.}
\label{fig:hamiltonianSSH}
\end{figure}
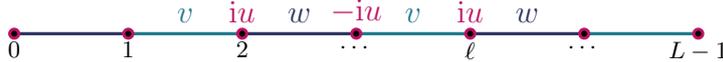

Before presenting our results, in this section we review the non-Hermitian model that we study in this paper.
We consider the non-Hermitian SSH (nH-SSH) chain with $PT$-symmetry on a discrete circle of $L = 2N$ sites, described by the Hamiltonian
\begin{equation}\label{eq:SSHhamiltonian}
    H = \sum_{j \in \ZZ_{N}} \lr{ - w \, c_{2j}^\dagger c_{2j+1} - v \, c_{2j-1}^\dagger c_{2j} + \mathrm{h.c.}}  + \ii u \sum_{j\in\ZZ_N} \lr{ c_{2j}^\dagger c_{2j} - c_{2j+1}^\dagger c_{2j+1} },
\end{equation}
with $u,v,w >0$. 
A schematic representation of this Hamiltonian is depicted in Fig. \ref{fig:hamiltonianSSH}.
We assume quasi-periodic boundary conditions, i.e., $c_{j+L} = e^{\ii \delta} c_j$, with $0 < \delta \ll 1$.
The reason for this choice will be explained later.
The model is a fermionic chain with nearest neighbours hoppings, which have alternating strength on even-odd links. The staggered imaginary chemical potential breaks the hermiticity of the Hamiltonian.
Notice that our conventions match those in Ref.~\cite{Tu:2021xje} after setting $v_1 = 0$ and $v_2 = v$ and identifying their up (down) sites with our even (odd) ones.

The Hamiltonian becomes block diagonal after a Fourier transform of the lattice operators, performed separately on the even and odd sites
\begin{align}
    \tilde c_{k,\mathrm e} &= \frac{1}{\sqrt N} \sum_{j\in \ZZ_N} e^{-\ii kj} c_{2j}\,, & 
    \tilde c_{k,\mathrm o} &= \frac{1}{\sqrt N} \sum_{j\in \ZZ_N} e^{-\ii kj} c_{2j+1}\,,
\end{align}
with 
\begin{equation}
    k \in \frac{2\pi}{N} \lr{\ZZ_N + \frac{\delta}{2\pi}},
\end{equation}
where the shift in momentum space is due to the $\delta$-twisted boundary conditions.
The Hamiltonian then becomes
\begin{equation}\label{eq:hamiltonian-fourier}
    H = \sum_k 
    \begin{pmatrix}
        \tilde c_{k,\mathrm e}^\dagger & \tilde c_{k,\mathrm o}^\dagger 
    \end{pmatrix}
    \begin{pmatrix}
        \ii u & - w- v e^{-\ii k} \\
        -w - ve^{\ii k} & -\ii u
    \end{pmatrix}
    \begin{pmatrix}
        \tilde c_{k,\mathrm e} \\ \tilde c_{k,\mathrm o} 
    \end{pmatrix},
\end{equation}
and the eigenvalues of the matrix in \cref{eq:hamiltonian-fourier} are the single particle energies.

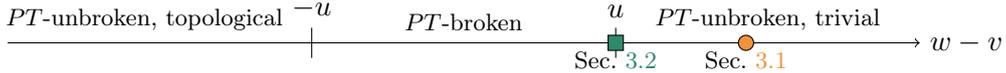
\begin{figure}
\centering
\begin{tikzpicture}
    \pgfmathsetmacro{\L}{12}
    \draw[->] (0,0) -- (\L,0) node[right]{$w-v$}; 
    \draw (\L/3,-0.2) -- (\L/3,0.2);
    \draw (2*\L/3,-0.2) -- (2*\L/3,0.2);
    \node[above] at (\L/3,0.2){$-u$};
    \node[above] at (2*\L/3,0.2){$u$};
    {
    \footnotesize
    \node[above] at (\L/6-0.2,0.05){$PT$-unbroken, topological};
    \node[above] at (\L/2,0.05){$PT$-broken};
    \node[above] at (\L*5/6,0.05){$PT$-unbroken, trivial};
    }
    
    \def \size {0.10}
    {
    \footnotesize
    \draw[fill=mycolor5,shift={(2*\L/3,0)}] (\size,\size) rectangle (-\size,-\size);
    \node[below] at (2*\L/3,0) {\hypersetup{allcolors=mycolor5}\cref{sec:EHgapless}};

    \draw[fill=mycolor4] (2*\L/3+\L/7,0) circle (\size);
    \node[below] at (2*\L/3+\L/7,0) {\hypersetup{allcolors=mycolor4}\cref{sec:EHgapped}};
    }
\end{tikzpicture}
\caption{Phase diagram of the nH-SSH model, explained in the main text. The orange circle and the green square mark the points in parameter space for which we study the entanglement Hamiltonian, reported in \cref{sec:EHgapped} and \cref{sec:EHgapless} respectively.}
\label{fig:phase-diagram}
\end{figure}

Varying the relative strengths of the parameters $u,v,w$, the model admits three different gapped phases~\cite{ChangRyu:2019}.
If $v-w \in (-u,u)$, the $PT$ symmetry is broken so that the energy spectrum is complex and the eigenvalues appear in complex conjugate pairs.
In the two phases $v-w > u$ or $v-w < -u$, the $PT$ symmetry is unbroken and the energy spectrum is real.
The latter two phases are distinguished by topological properties, as discussed in~\cite{Lieu:2018topoSSH}.
The resulting phase diagram is given in \cref{fig:phase-diagram}.

Two critical points occur for $v-w = \pm u$. In these cases, the single particle spectrum is $\epsilon_{\pm,k} = \pm \sqrt{2vw(1+\cos k)}$ and the gap closes at $k=\pi$, leading locally to a linear spectrum with speed of sound
\begin{equation}\label{eq:sound-velocity}
    c_S = \sqrt{vw}.
\end{equation}
Moreover, at $k=\pi$ the kernel of the Hamiltonian \eqref{eq:hamiltonian-fourier} is not diagonalisable, as it is made of a $2\times 2$ Jordan block.
This is called an exceptional point in momentum space. 
The exceptional point occurs because, as $k \to \pi$, the two eigenspaces become more and more collinear, and they perfectly coincide at $k=\pi$. 

Finally, since the Hamiltonian is a linear combination of terms of the form $c^\dagger_i c_j$, it is invariant under the $U(1)$ generated by
\begin{equation}
    Q = \sum_{j \in \ZZ_{2N}} c^\dagger_j c_j\,.
\label{QQ}    
\end{equation}

In this paper, we will investigate the ground-state of the system in the $PT$-unbroken trivial phase and the critical point between the $PT$-unbroken trivial phase and the $PT$-broken phase, marked in \cref{fig:phase-diagram} with a orange circle and a green square, respectively.
In~\cite{ChangRyu:2019}, the latter point has been identified with the fermionic $bc$-ghost CFT with central charge $c=-2$, which we review in the following section.

\subsection{\texorpdfstring{$bc$}{bc}-ghost CFT}

The $bc$-ghost conformal field theories are a family of CFTs governed by the following action~\cite{FriedanMartinecShenker:1985ge, Guruswamy:1996rk, Kausch:1995py, Kausch:2000fu, DiFrancesco:1997nk}
\begin{equation}
    S = \int \dd^2 z \lr{ b \: \bar \partial c + \bar b \, \partial \bar c },
\end{equation}
where $b$ and $c$ are anticommuting holomorphic fields and $\bar b$ and $\bar c$ are the corresponding anti-holomorphic fields.
The different members of this family are distinguished by the value of the central charge and by the conformal dimension of the fields $c$ and $b$.
In particular, the CFT which describes the nH-SSH critical point is the one with central charge $c= -2$~\cite{ChangRyu:2019}, in which the fields have conformal weight $h_b = 1, h_c = 0$.
All these theories have a conserved current $J =\, \nord{cb}$ so that the field $c$ has charge $1$ and $b$ has charge $-1$, independently of the specific realisation and central charge.

The conformal field theory with $c=-2$ is one of the simplest instances of a logarithmic CFT \cite{Kausch:1995py}, incorporating reducible but not indecomposable representations of the Virasoro algebra.
Specifically, the fields $c$ and the identity field share the same conformal weights, leading to the formation of a $2$-dimensional Jordan block in the Virasoro modes $L_0$ and $\bar L_0$.
This phenomenon occurs exclusively in the untwisted sector of the theory, which corresponds to periodic boundary conditions on a cylinder.
In the scenario where $\delta$-twisted boundary conditions are adopted, the fields acquire a phase factor $e^{\ii 2\pi \delta}$ as they move around the non-contractible loop of the cylinder.
Consequently, the identity field is no longer part of the spectrum, and the system's ground state becomes associated with the twist field $\sigma_\delta$~\cite{Kausch:1995py}.
The conformal dimension of $\sigma_\delta$ is given by $h_{\sigma_\delta} = \delta(\delta-1)/2$, which is negative for $\delta \neq 0$.
This implies that for $\delta \neq 0$, there is no Jordan block for $L_0$ and $\bar L_0$, effectively eliminating the logarithmic singularities.
It is noteworthy that the presence of the Jordan block in periodic boundary conditions and its absence in the twisted sectors draws a further analogy with the nH-SSH model.

\subsection{Left-right ground-state}\label{sec:lrstate}
Before concluding this brief review, we would like to emphasise the states that are the focus of this paper.
First, in both of the cases we consider (see \cref{fig:phase-diagram}), the Hamiltonian has a real spectrum, thus there is a well defined notion of a ground state as the eigenstate with minimum energy eigenvalue.
We denote by $\ket{R}$ the right ground state of the Hamiltonian, defined by $H \ket{R} = E_{gs} \ket{R}$, while we denote with $\bra{L}$ the left ground-state, defined by $\bra{L} H = E_{gs} \bra{L}$.
Since the Hamiltonian is non-Hermitian, the left ground state is not the ``bra'' of the right ground state, in other words, $\ket{L} \neq \ket{R}$.

We consider the density matrix $\rho = \ket{R} \bra{L}$, which we call the left-right ground state \cite{Brody:2013axr, Couvreur:2016mbr, Dupic:2017hpb, Herviou:2019, ChangRyu:2019, Tu:2021xje, Fossati:2023zyz, Tang:2023akr}.
Indeed, this can be seen as the zero-temperature limit of the thermal state $e^{-\beta H}/Z$ and therefore is the most natural object to be studied in field theory.
The density matrix $\rho$ is positive semi-definite but not Hermitian and therefore the reduced density matrix $\rho_A$ is not positive semi-definite.
This means that the entanglement entropy between a subsystem and its complement can be negative.
Indeed,  the entanglement entropy scales as $c/3 \log \ell$, with $c=-2$~\cite{ChangRyu:2019}. 

The symmetry-resolved entanglement, relative to the $U(1)$ symmetry \eqref{QQ}, at the critical point has been studied in~\cite{Fossati:2023zyz}. 
Of relevance for this paper, it has been understood that the eigenvalues of the reduced density matrix are either positive or negative depending on the sign of the charge sector, namely $\operatorname{sign} \lambda_q = (-1)^{q- \langle Q_A \rangle}$, where  $\lambda_q$ stands for an eigenvalue of $\rho_A$ in the charge sector $q$ of $Q_A$ (i.e. the charge \eqref{QQ} restricted to $A$).
We will show in \cref{sec:EHgapless} that we can identify the source of this behaviour in the form of the entanglement Hamiltonian.

\subsection{Correlation function}

A key object in the analysis of the entanglement Hamiltonian of the left-right
ground state is the two-point correlation matrix $C$ with entries \cite{ChangRyu:2019, Fossati:2023zyz}
\begin{equation}
\label{eq:correl_matrix_2}
    C_{2j+a,2l+b} = \langle L| c_{2j+a}^\dagger c_{2l+b} |R\rangle = \frac{1}{N} \sum_k e^{-\ii k(j-l)} \mathcal{G}(k)_{ab} \, , \quad a,b \in \{ 0, 1\} ,
\end{equation}
with
\begin{equation}
    \mathcal{G}(k)=\frac{1}{2}
    \begin{pmatrix}
        1-\cos(2\xi_k) & -\sqrt{\frac{\eta_k^*}{\eta_k}}\sin(2\xi_k) \\
        -\sqrt{\frac{\eta_k}{\eta_k^*}}\sin(2\xi_k) & 1+\cos(2\xi_k)
    \end{pmatrix},
\end{equation}
where $2 \xi_k = \tan^{-1} ( \abs{\eta_k} / (i u) )$, $\eta_k = - w - v e^{-ik}$.
Due to the dimerization of the hopping amplitudes $v$, $w$, the correlation matrix $C$ presents a block structure.
In the thermodynamic limit $L\to\infty$, $C$ is a block Toeplitz matrix generated by the symbol $\mathcal{G}$.

\section{Lattice entanglement Hamiltonians of the non-Hermitian SSH model}\label{sec:EH}

This section contains the main results of this paper, the numerical lattice entanglement Hamiltonian in the non-Hermitian SSH model and an analytic conjecture for its behaviour.
In order to compute numerically the lattice EH we use the known relation between fermionic Gaussian states and the correlation matrix.
Notice first that since the Hamiltonian~\eqref{eq:SSHhamiltonian} is quadratic, the ground state is Gaussian~\cite{ChangRyu:2019, Fossati:2023zyz} and the reduced density matrix can be written as 
\begin{equation}\label{eq:latticeEH}
    \rho_A = \frac{1}{Z_A} \exp\!\left \{ - \sum_{i,j\in A} c_{i}^\dagger k^A_{ij} c_{j} \right \},
\end{equation}
where $k^A_{ij}$ is the kernel of the EH, i.e., the single particle entanglement Hamiltonian.
For Gaussian states as in \cref{eq:latticeEH}, the kernel $k^A$ can be obtained from the knowledge of the reduced correlation matrix, i.e. the matrix \eqref{eq:correl_matrix_2} with indexes restricted to $A$,  $\left ( C_A \right )_{ij} = \left ( C \right )_{i, j \in A}$. Using Peschel's formula~\cite{Chung:2001zz, Peschel:2002yqj, Eisler:2009vye} one has
\begin{equation}\label{eq:peschels}
    k^A = \log\left [ C_A^{-1} - \mathbb{I} \right ]^T,
\end{equation}
where $T$ denotes the matrix transpose.
While \cref{eq:peschels} was initially derived for Hermitian models, as discussed in Refs.~\cite{ChangRyu:2019, Fossati:2023zyz}, it remains valid in the non-Hermitian one under consideration.
In Refs.~\cite{ChangRyu:2019, Fossati:2023zyz} the restricted correlation matrix of the non-Hermitian SSH model is used for the computation of the entanglement spectrum and the entropies. 
In the following we will compute the kernel of the entanglement Hamiltonian using the correlation matrix \eqref{eq:correl_matrix_2}.

We remark that the numerical computation of the formula  \eqref{eq:peschels} suffers from numerical instabilities and must be conducted at high precision.
The reason for this instability is that many eigenvalues of the correlation matrix $C_A$ are arbitrarily close to $0$ and $1$, and as a consequence the matrix inside of the logarithm in \cref{eq:peschels} has eigenvalues which are very close to $0$ or very large.
In our study we used the \verb|python| library \verb|mpmath|~\cite{mpmath} and the software \verb|Mathematica|, keeping up to 500 digits.

In rest of this section, we present the results for the entanglement Hamiltonian of an interval $A = [0, \ell]$ in the left-right ground state.
We first study the topologically trivial gapped phase $w-v > u$ with periodic boundary conditions and we compare with the known results in unitary integrable lattice models~\cite{Eisler:2020lyn}.
We then consider the critical point $w-v = u$ with a small twisting of the boundary conditions $\delta = 10^{-7}$, which as we explained in \cref{sec:model} is described by the $c = -2$ $bc$-ghost CFT.
We compare the results with the continuum prediction from unitary CFTs and we use our observations to formulate a conjecture for the entanglement Hamiltonian of an interval in the ground state of the $bc$-ghost theory.

\subsection{Entanglement Hamiltonian in the trivial gapped phase}\label{sec:EHgapped}

\begin{figure}[t]
    \centering
    \begin{subfigure}[b]{0.49\textwidth}
        \centering
        \includegraphics[width=\textwidth]{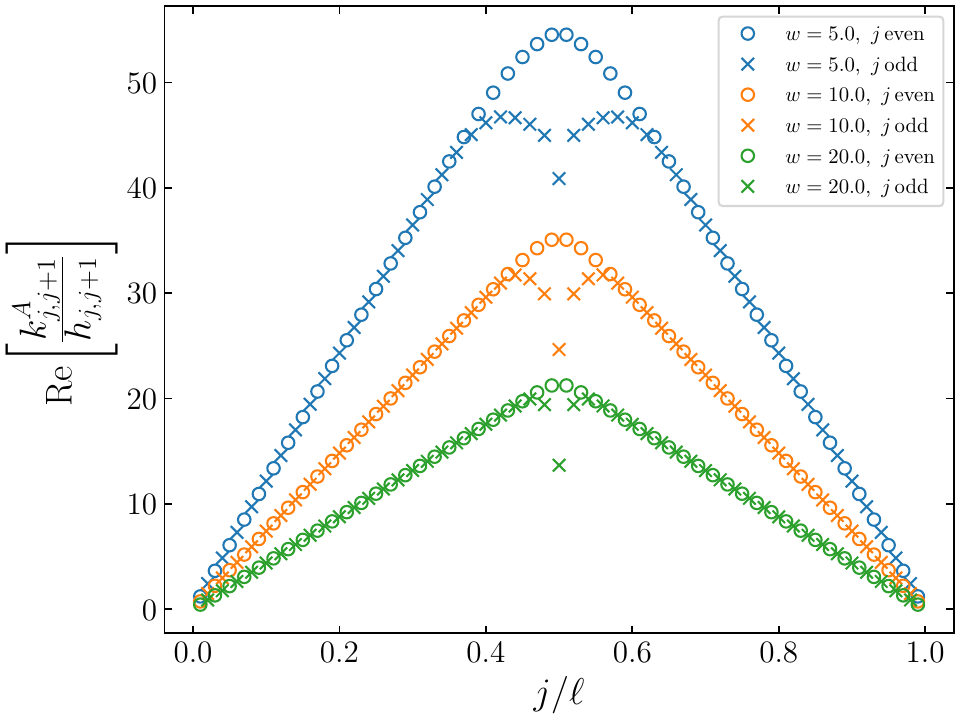}
        \caption{Real part of the nearest-neighbour coupling $k^A_{j, j+1}$.}
        \label{fig:gappedEHreal}
    \end{subfigure}
    \hfill
    \begin{subfigure}[b]{0.49\textwidth}
        \centering
        \includegraphics[width=\textwidth]{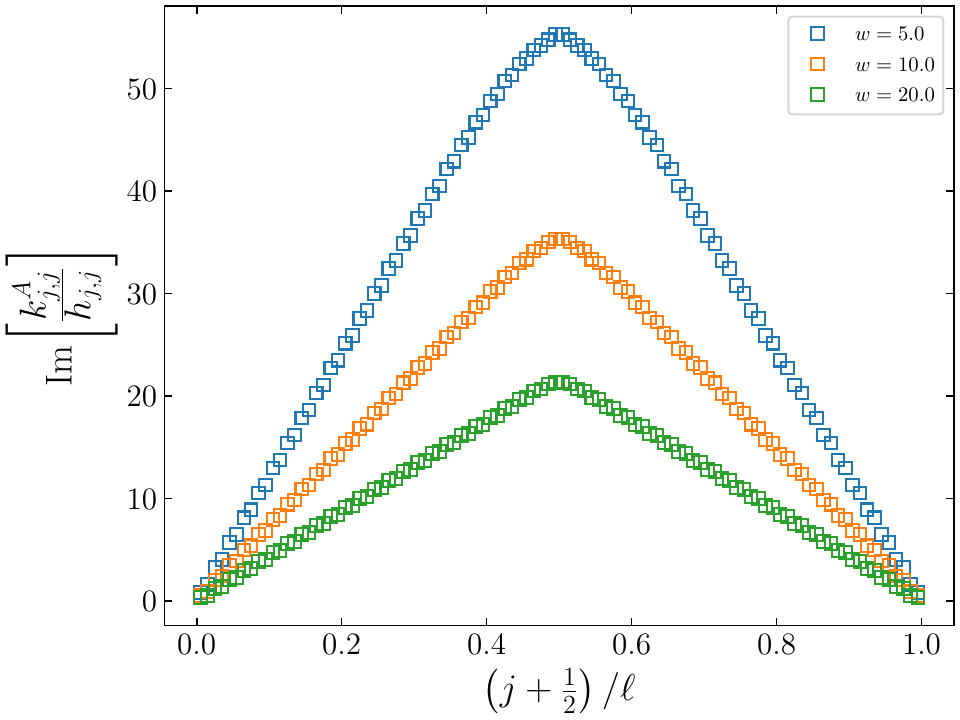}
        \caption{Imaginary part of the chemical potential $k^A_{j, j}$.}
        \label{fig:gappedEHimag}
    \end{subfigure}
    \caption{Entanglement temperature in the gapped phase $w-v>u$. In both plots we fix $v = u = 1$ and we consider different values of $w = 5, 10$ and $20$ and we take a subsystem of length $\ell = 100$ in a full system of total length $L = 2000$. 
    In the left plot we report, as a function of $j/\ell$, the real part of the ratio of the nearest-neighbour EH coupling $k^A_{j, j+1}$ with the coupling ($h_{j,j+1}$), i.e. $-w$ for even $j$ (circles) and $-v$ for odd $j$ (crosses). 
    The purpose of this ratio is to isolate the entanglement temperature. 
    Apart from a small region in the center of the interval, the ratio follows the expected triangular shape (see discussion below \cref{eq:TTISSH}). In the right plot we report the imaginary part of the ratio between the staggered imaginary chemical potential $k^A_{j, j}$ with $+ u$ ($-u$) for even (odd) site $j$. Again, up to a small finite size oscillation, the ratio follows the predicted triangular shape.}
    \label{fig:gappedEH}
\end{figure}

\begin{figure}[t]
    \centering
    \begin{subfigure}[b]{0.49\textwidth}
        \centering
        \includegraphics[width=\textwidth]{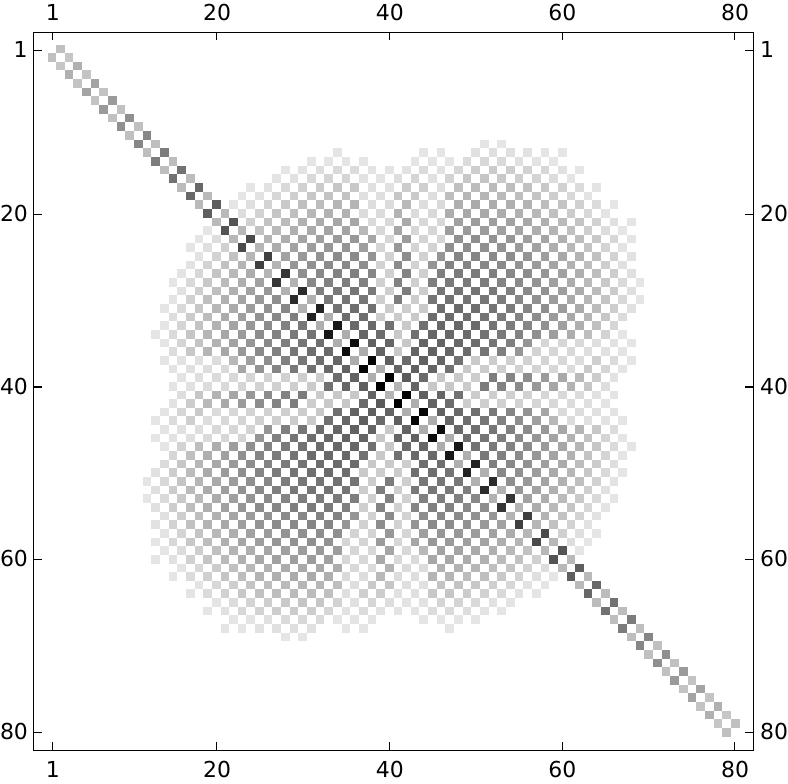}
        \caption{Real part of the EH kernel $k^A$.}
        \label{fig:matrixplotGappedreal}
    \end{subfigure}
    \hfill
    \begin{subfigure}[b]{0.49\textwidth}
        \centering
        \includegraphics[width=\textwidth]{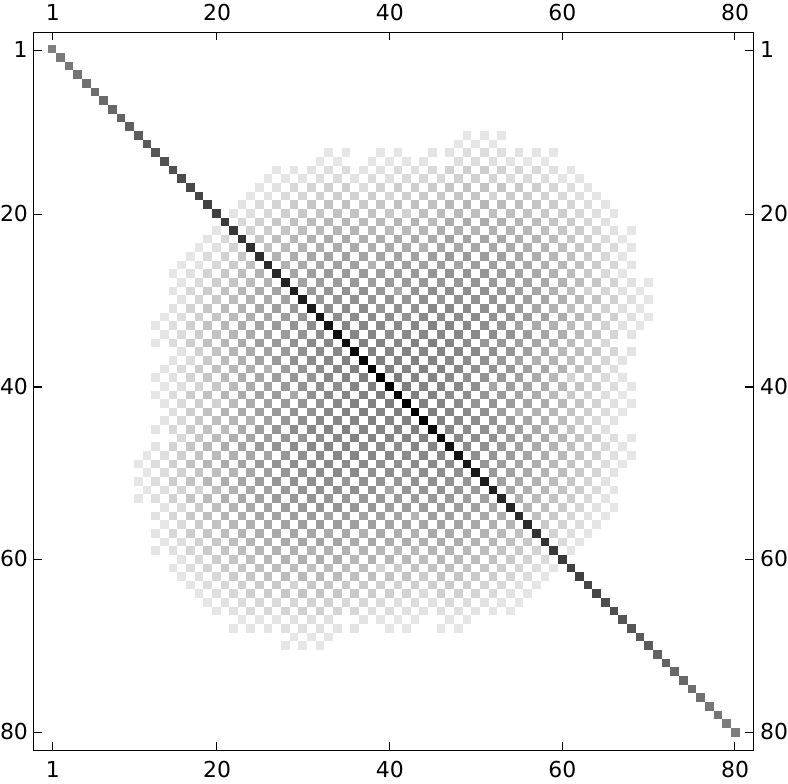}
        \caption{Imaginary part of the EH kernel $k^A$.}
        \label{fig:matrixplotGappedimag}
    \end{subfigure}
    \caption{Matrix plot of the entanglement Hamiltonian kernel $k^A$ in the gapped phase $w-v > u$ with $w = 20$, $v = 2$ and $u = 2$, for an interval of length $\ell=80$ in a system of length $L = 2000$. 
    Left (Right): Absolute value of the real (imaginary) part 
    of $k^A$. 
    Consistently with the Tetel'man-Thacker behaviour~\eqref{eq:TTISSH}, near the two endpoints the only non-vanishing elements of the entanglement Hamiltonian are the imaginary chemical potential (main diagonal in the right plot) and the coupling between nearest-neighbours (first sub-diagonals in the left plots). 
    The latter couplings (left) display the alternating value between the odd and even sites (see \cref{eq:TTISSH}). In the middle of the interval, the entanglement Hamiltonian deviates from \cref{eq:TTISSH} and also couplings at higher distances are non-zero.}
    \label{fig:matrixplotGapped}
\end{figure}

Before studying the non-Hermitian model, it is instructive to first recall the known results in unitary gapped lattice models, in order to compare them with ours.
As we reported in \cref{eq:TTI} in \cref{sec:intro}, in certain integrable models the entanglement Hamiltonian in the half-space follows the structure recognised by Tetel'man, Itoyama and Thacker, i.e., the EH is proportional to the Hamiltonian density with a local temperature equal to the lattice site, analogous to a lattice Bisognano-Wichmann behaviour~\cite{Baxter:1976uh, Baxter:1977ui, Tetelman, Thacker:1985gz, Itoyama:1986ad, Dalmonte:2022rlo}.
If we instead consider a finite interval, in the general case there are very few known analytic results.
If the gap is sufficiently large, however, in Ref.~\cite{Eisler:2020lyn} it was observed via numerical computations that near the two endpoints of the interval the EH follows the half-space result of \cref{eq:TTI}, only deviating from this behaviour in the middle of the interval, which give rise to a characteristic triangular entanglement temperature.
This triangular behaviour has been observed in several Hermitian models, such as the Hermitian Su-Schrieffer-Heeger model (or dimerised hopping chain) and the harmonic chain~\cite{Eisler:2020lyn}.
It is arguably a straightforward consequence of cluster decomposition, which is independent of unitarity.  
It is therefore natural to wonder if this factorisation holds also for the non-Hermitian model under study.
Another important consequence of \cref{eq:TTI} is that in unitary lattice integrable modes, the half-space lattice EH does not couple fermions at distances larger than those in the corresponding lattice Hamiltonian.
Correspondingly, within an interval, it was noted that near the endpoints, the entanglement Hamiltonian does not exhibit higher couplings, only manifesting them in the crossover region at the center~\cite{Eisler:2020lyn}.

Let us now consider the non-Hermitian SSH model.
Assuming that the structure of the entanglement Hamiltonian in \cref{eq:TTI} holds also for this theory, from the Hamiltonian in \cref{eq:SSHhamiltonian} we can conjecture that the half-space lattice EH takes the form
\begin{equation}\label{eq:TTISSH}\begin{split}
    K_A \propto \sum_{j = 0}^{+\infty} \bigg [ & \left (2j \right ) w \left ( c^\dagger_{2j} c_{2j+1} + c^\dagger_{2j+1}  c_{2j} \right ) +  \left (2j + 1 \right ) v \left ( c^\dagger_{2j-1} c_{2j} + c^\dagger_{2j}  c_{2j-1} \right ) +\\
    &\hspace{3cm} + \ii \left ( 2j + \frac{1}{2} \right ) u \, c^\dagger_{2j} c_{2j} -\ii \left ( 2j + \frac{3}{2} \right ) u\, c^\dagger_{2j+1}  c_{2j+1} \bigg ],
\end{split}\end{equation}
with some unknown proportionality constant.
Since we cannot access numerically the full EH of the half-space, in order to test the conjecture in \cref{eq:TTISSH} we study the EH of an interval $[0, \ell]$ in a finite system of length $L \gg \ell$.
In analogy with the unitary case, we expect that for a sufficiently large gap, near the endpoints the entanglement Hamiltonian will follow the half-space result in \cref{eq:TTISSH}, with a crossover in the middle of the interval, giving rise to the typical triangular shape.

In \cref{fig:gappedEH} we report the results of the numerical calculation of the lattice entanglement Hamiltonian in the gapped phase, for an interval of length $\ell = 100$ in a system of total length $L = 2000$ with periodic boundary conditions.
We fix the parameters $v = u = 1$ and we study different gaps by varying the value of $w$, in particular we take $w = 5, 10$ and $20$.
The plots report the ratio between the kernel of the EH, $k^A$, obtained from \cref{eq:peschels} and the one of the Hamiltonian $h$ in \cref{eq:SSHhamiltonian} as a function of the lattice site.
On the left, in \cref{fig:gappedEHimag} we report the real part of the nearest-neighbour coupling $k^A_{j, j+1}$, divided by $(-w)$ for $j$ even (circles) and by $(-v)$ for $j$ odd (crosses).
Dividing by these coupling constants, we isolate the entanglement temperature, which is expected to follow the triangular shape (see \cref{eq:TTISSH} and discussion below).
Indeed we see that, apart from a small crossover region in the center of the interval, the nearest-neighbour coupling follows the expected behaviour for all values of $w$ that we considered.
This behaviour is completely analogous to what observed in Ref.~\cite{Eisler:2020lyn} for the dimerised hopping chain.
The novel result is reported in the right plot, in \cref{fig:gappedEHimag}, where we show the staggered imaginary chemical potential $k^A_{j, j}$, divided by $u$ for $j$ even and by $(-u)$ for $j$ odd.
Again, the role of this division is to isolate the entanglement temperature, which should agree with the one obtained from the nearest-neighbour coupling.
Indeed we observe that, apart from a small oscillation due to finite size effects, the imaginary chemical potential follows the same triangular shape as the nearest-neighbour coupling, as expected from our conjecture in \cref{eq:TTISSH}.

As a further check, in \cref{fig:matrixplotGapped} we report the matrix plots of the real (left plot) and of the imaginary parts (right plot) of the single particle EH $k^A$.
According to our conjecture in \cref{eq:TTISSH}, the half-space EH does not couple fermions at distances higher than one, similarly to what happens for unitary integrable models in \cref{eq:TTI}.
In the left plot in \cref{fig:matrixplotGappedreal}, we see that near the endpoints the only non-zero elements of the real part of the EH kernel are the nearest-neighbour couplings $k^A_{j, j+1}$ and $k^A_{j, j-1}$.
The higher couplings are non-zero only in a crossover region in the middle of the interval, as expected.
This behaviour is again completely analogous to what was observed in Ref.~\cite{Eisler:2020lyn} for the dimerised hopping chain.
The new results are given by the imaginary part, shown in the right plot in \cref{fig:matrixplotGappedimag}.
We see that also the imaginary part follows the expected behaviour, with only the main diagonal $k^A$ being significantly different from zero near the endpoints.
This confirms the validity of our local conjecture in \cref{eq:TTISSH} for the half space EH in the non-Hermitian SSH model. 
We remark that this is the first observation of a Bisognano-Wichmann like behaviour in a non-Hermitian model.

Before concluding this section, we wish to comment on the proportionality constant in \cref{eq:TTISSH}, i.e., the slope of the triangles in \cref{fig:gappedEH}.
This constant is actually related to the velocity of the excitations in the gapped model.
In Ref.~\cite{Eisler:2020lyn}, the analogous proportionality constant in the dimerised hopping chain was computed analytically using the knowledge of the exact CTM.
It would be interesting to obtain analytically the CTM in the non-Hermitian SSH model, which would refine our conjecture~\eqref{eq:TTISSH} for the half-space EH.
This computation would not only allow us to predict the slope of the linearly increasing entanglement temperature, but it could also provide a quantitative understanding of the finite size oscillations of the chemical potential in \cref{fig:gappedEHimag} which are not captured by \cref{eq:TTISSH}.
This is however a rather involved calculation which goes beyond the scope of this work.

\subsection{Entanglement Hamiltonian at the critical point}\label{sec:EHgapless}

\begin{figure}[t]
    \centering
    \includegraphics[width=0.7\textwidth]{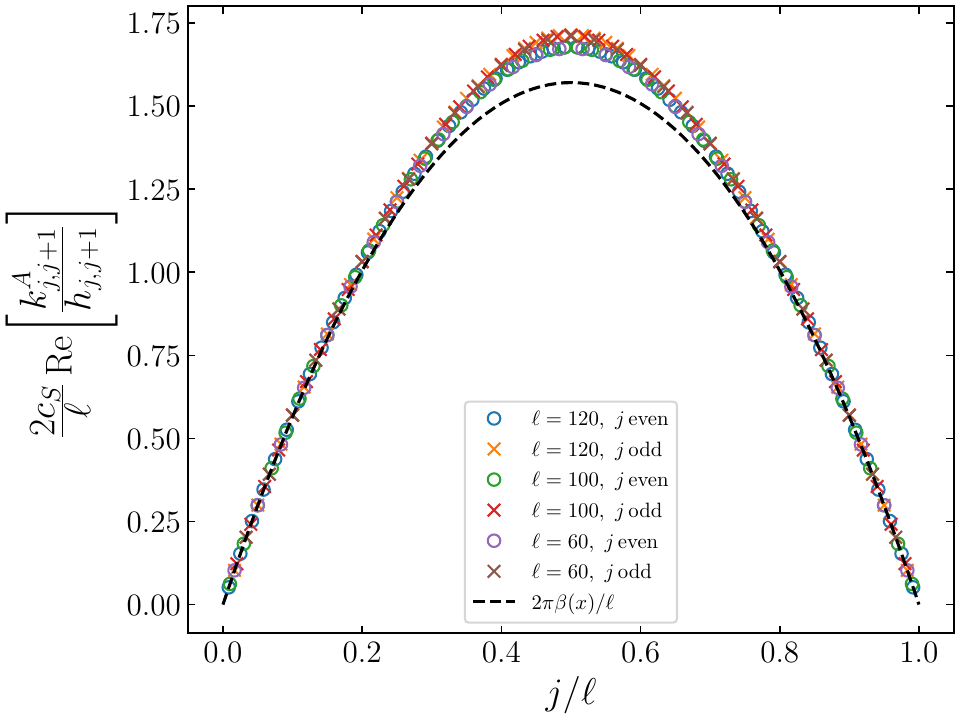}
    \caption{Real part of the ratio of the EH nearest-neighbour coupling $k^A_{j, j+1}$ with the coupling $w$ ($v$) for $j$ even (odd), rescaled by $2 c_S / \ell$, where $\ell$ is the length of the interval and $c_S$ is the speed of sound~\eqref{eq:sound-velocity}. The circles represent even sites and the crosses are odd sites. For all lengths considered we observe a perfect collapse. The black dashed parabola is the field theory prediction for the local temperature $2\pi \beta(x)$ in \cref{eq:CFTinterval}, divided by $\ell$. Near the endpoints of the interval we find a very good agreement between the lattice result and the field theory. The deviation in the middle of the interval is due to the contribution of higher couplings, analogously to what happens in Hermitian lattice models.}
    \label{fig:gaplessEHreal}
\end{figure}

\begin{figure}[t]
    \centering
    \includegraphics[width=0.7\textwidth]{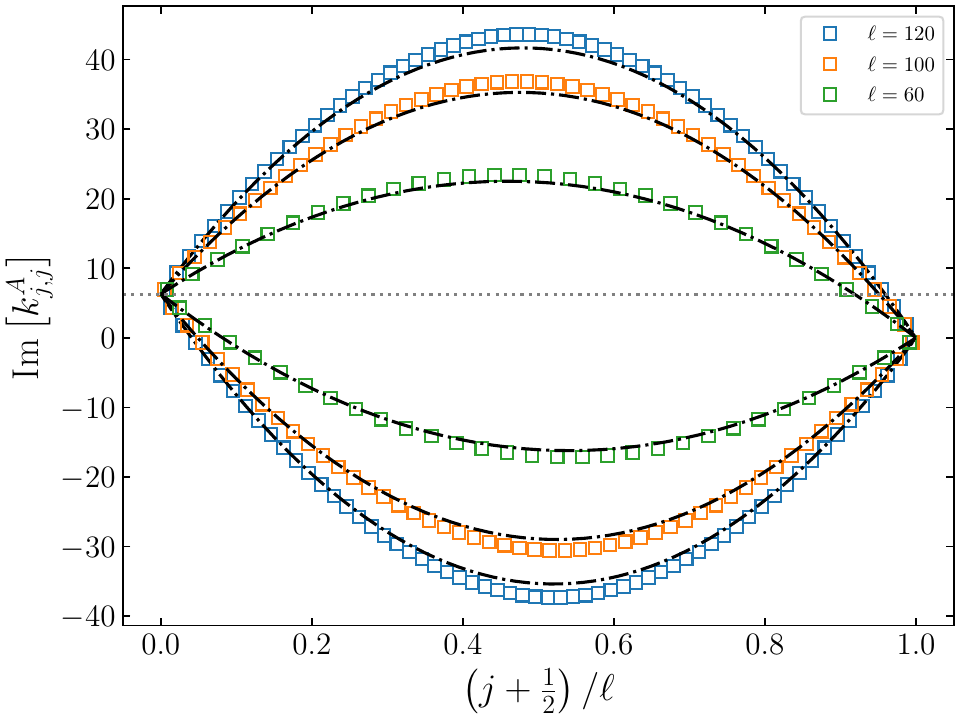}
    \caption{Imaginary chemical potential $k^A_{j,j}$ at criticality $w-v=u$ for different lengths of the interval $\ell = 60, 100$ and $120$. The black dash-dotted curves are reported in \cref{eq:curve} and are obtained as the sum of the naive field theory prediction for the entanglement temperature in \cref{eq:CFTinterval} and of the conjectured form of the novel term in \cref{eq:conjecturelattice}. 
    Close to the endpoints we observe a perfect agreement which becomes slightly worse in the middle of the interval.}
    \label{fig:gaplessEHimag}
\end{figure}

\begin{figure}[t]
    \centering
    \includegraphics[width=0.7\textwidth]{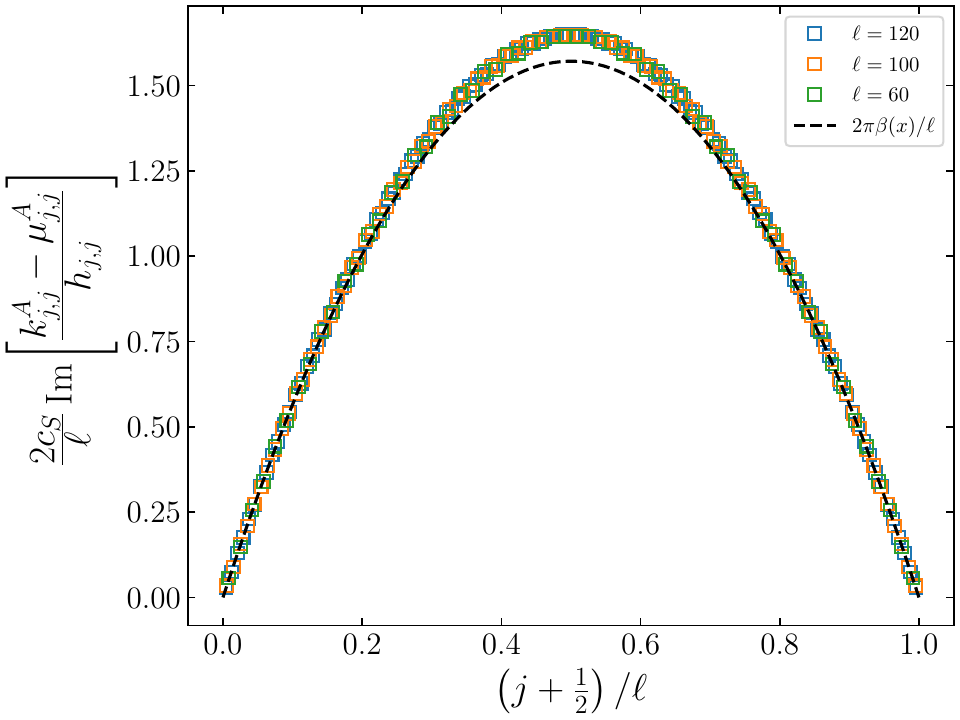}
    \caption{Imaginary part of $(k^A_{j,j} - \mu^A_{j,j})/h_{j,j}$ (i.e. the difference between the EH the chemical potential and $\mu^A_{j,j}$ in \cref{eq:conjecturelattice}, all in units of $h_{j,j}$), rescaled with $2 c_S / \ell$, where $c_S$ is the speed of sound in \cref{eq:sound-velocity}. We consider intervals of length $\ell = 60, 100$ and $120$ in a system of total length $L = 2000$, with parameters $w = 1.5$, $v = 1$ and $u = w-v = 0.5$. 
    For all $\ell$, we observe a perfect collapse, suggesting that we have successfully isolated the scaling part. The black dashed curve is the CFT prediction for the entanglement temperature $2\pi\beta$ in \cref{eq:CFTinterval} divided by $\ell$. Analogously to the nearest-neighbour coupling in \cref{fig:gaplessEHreal}, the agreement is perfect at the endpoints and is slightly worse in the middle of the interval, due to the contribution of higher order couplings.}
    \label{fig:gaplessEHimagMinusLine}
\end{figure}

In this section we study the EH at the critical point $w-v=u$ (green square in \cref{fig:phase-diagram}).
As discussed in \cref{sec:model}, at the critical point and for periodic boundary conditions, the lattice Hamiltonian~\eqref{eq:SSHhamiltonian} presents a Jordan block. Then, to treat the system numerically we need to introduce a small twisting of the boundary conditions $\delta$~\cite{ChangRyu:2019, Fossati:2023zyz}.
In all the following discussion we fix $\delta = 10^{-7}$.
In full analogy to the study we performed for the gapped phase in \cref{sec:EHgapped}, we compute numerically the lattice EH kernel $k^A$ using \cref{eq:peschels}, performing all calculations at high precision.
However, at the critical point there is an additional subtlety.
In Ref.~\cite{ChangRyu:2019} it was shown that at criticality all eigenvalues $\nu_j$ of the correlation matrix are real and lie outside of the interval $[0, 1]$.
As a consequence, the matrix appearing inside the logarithm in \cref{eq:peschels} has all negative eigenvalues (see also Ref.~\cite{Fossati:2023zyz}).
This is susceptible to numerical instabilities, giving an imaginary part of the logarithm which (unphysically) oscillates wildly between $+\ii\pi$ and $-\ii\pi$.
In this work we always fix it to be equal to $+\ii \pi$.

Before presenting our numerical results for the critical non-Hermitian SSH model, we would like to reiterate what occurs in the case of unitary gapless models.
According to \cref{eq:CFTinterval}, in the vacuum of the CFT describing the continuum limit of a critical model, the entanglement Hamiltonian of an interval is proportional to the energy density with a parabolic entanglement temperature $\beta(x)$~\cite{Hislop:1981uh, Casini:2011kv, Wong:2013gua, Cardy:2016fqc}.
One could be tempted to conclude that the EH of an interval in a critical model should be proportional to the critical Hamiltonian density with the parabolic temperature in \cref{eq:CFTinterval}.
This behaviour would in particular imply that all terms of the lattice EH that couple fermions at distances higher than those in the Hamiltonian must be negligible.
However, as recognised first in Ref.~\cite{Arias:2016nip}, this is not the case, and the EH contains couplings at arbitrary distances (see also~\cite{Giudici:2018izb, Mendes-Santos:2019tmf, Zhang:2020mjv}).
Moreover, when expanding the lattice fermions in the lattice spacing, all these higher couplings contribute to the continuum energy density $T_{00}$~\cite{Eisler:2019rnr, Eisler:2022rnp}.
In Refs.~\cite{Eisler:2019rnr, Eisler:2022rnp, DiGiulio:2019cxv} it was shown that in order to recover the CFT entanglement temperature $\beta(x)$ in \cref{eq:CFTinterval}, it is necessary to perform a careful continuum limit which takes into account all of these higher contributions.
This limiting procedure has allowed to reconstruct the CFT entanglement Hamiltonian in many systems at criticality, both at finite temperature and in the ground state~\cite{Eisler:2019rnr, DiGiulio:2019cxv, Eisler:2022rnp} and also in the presence of boundaries~\cite{DiGiulio:2019cxv, Eisler:2022rnp, Rottoli:2022plr}, in inhomogeneous and out-of-equilibrium systems~\cite{Rottoli:2022ego} and in higher dimensions~\cite{Javerzat:2021hxt}.
In Refs.~\cite{Rottoli:2022plr, Rottoli:2023xov} it has also been extended to the recently introduced negativity Hamiltonian~\cite{Murciano:2022vhe}, i.e., the logarithm of the partial transposed density matrix.
On the other hand, this limit is highly dependent on the lattice model and, to date, it is only understood in the case of free massless lattice fermions and the harmonic chain.

Considering now the non-Hermitian SSH model, in \cref{fig:gaplessEHreal,fig:gaplessEHimag}, we report the numerical lattice EH, obtained from \cref{eq:peschels} with a choice of parameters $w = 1.5$, $v = 1$ and $u = w-v = 0.5$ and different interval lengths $\ell = 60, 100$ and $120$ in a total system of length $L = 2000$.
In \cref{fig:gaplessEHreal} we plot the real part of the nearest-neighbour coupling $k^A_{j, j+1}$, divided by $(-w)$ for $j$ even and by $(-v)$ for $j$ odd, analogously to what we have done in the massive case.
We further make the quantity dimensionless by multiplying it by $2 c_S / \ell$, where $c_S$ is the speed of sound~\eqref{eq:sound-velocity} in the critical lattice model.
Indeed, notice that if we reintroduce the dimensions, $k^A$ is dimensionless, while $w$ and $v$ have the dimensions of an inverse time. 
We observe a perfect collapse for all the lengths considered.
The black dashed line in \cref{fig:gaplessEHreal} is the parabolic entanglement temperature $2\pi\beta(x)$ for unitary CFTs reported in \cref{eq:CFTinterval}, divided by the length of the interval $\ell$.
While near the endpoints we find a good agreement, we see a deviation in the middle of the interval.
Similarly to what happens for unitary lattice models, the origin of this discrepancy is the presence of higher couplings which in the continuum limit give contributions to the continuum energy density.
We expect that a proper continuum limit should exactly reproduce the parabola in \cref{eq:CFTinterval} (as for Hermitian free fermions~\cite{Eisler:2019rnr}), but this is beyond our goals.

In \cref{fig:gaplessEHimag} we instead report the staggered imaginary chemical potential (the alternating sign with respect to \cref{fig:gappedEHimag} is due to not having divided by either $u$ or $(-u)$).
This quantity displays the most significant difference with respect to the Hermitian case. 
For all the lengths $\ell$ of the interval, at the left endpoint $j/\ell = 0$ the chemical potential takes the value $2\pi \ii$ (grey dotted line), while at the right one $j /\ell = 1$ it vanishes.
Based on this observation, we conjecture that besides the approximate parabolic result, at the critical point appears an additional term of the form 
\begin{equation}\label{eq:conjecturelattice}
    \sum_{j=0}^{\ell} \mu^A_{j, j} \, c^\dagger_{j} c_{j} = 2\pi\ii\, \sum_{j=0}^{\ell} \lr{1 + \frac{\lr{j+\frac{1}{2}}}{\ell}} c^\dagger_{j} c_{j},
\end{equation}
i.e., a chemical potential term which interpolates linearly between $2\pi\ii$ and $0$.
We remark that, differently from the parabolic entanglement temperature $\beta(x)$ in \cref{eq:CFTinterval}, this novel term does not scale with the system size.
In order to check \cref{eq:conjecturelattice}, in \cref{fig:gaplessEHimag} we compare the two curves (dash-dotted black lines)
\begin{equation}\label{eq:curve}
    \frac{\pi\lr{\pm u}}{c_S} \lr{\frac{(\ell -x) x}{\ell}} + 2\pi \lr{1-\frac{x}{\ell}},
\end{equation}
with the imaginary part of the EH chemical potential term for $\ell = 60, 100$ and $120$.
Near the endpoints we find a perfect match for all the lengths considered, while the agreement gets slightly worse in the middle of the interval, but still  acceptable

To facilitate the comparison, we extract the part of the EH chemical potential that scales with the length of the interval by subtracting the conjectured form $\mu^A_{j, j}$ in \cref{eq:conjecturelattice} from the numerical result for $k^A_{j,j}$.
We then divide by $u$ for $j$ even and by $(-u)$ for $j$ odd to isolate the entanglement temperature and we rescale with $2c_S/\ell$ to make the quantity dimensionless.
For all the values of the length considered we observe a perfect collapse, which suggests that the novel non-scaling term $\mu^A_{j, j}$ takes indeed the conjectured form  \eqref{eq:conjecturelattice}.
The black dashed curve is again the parabolic CFT prediction for the entanglement temperature in \cref{eq:CFTinterval} divided by $\ell$.
Once again, we have a perfect agreement near the endpoints of the interval, while we observe a deviation in the middle.
This deviation is always due to the presence of contributions from higher couplings.

Summing up our finding, recalling from \cref{sec:model} that the critical point is described by the $c = -2$ $bc$-ghost CFT, we propose that the continuum limit of the difference $(k^A - \mu^A)$ must reproduce the continuum CFT entanglement Hamiltonian in \cref{eq:CFTinterval}.
Meanwhile, the continuum limit associated with the new chemical potential term $\mu^A$ in \cref{eq:conjecturelattice} will yield
\begin{equation}\label{eq:continuumconjecture}
    \sum_{j=0}^{\ell} \mu^A_{j, j}\, c_j ^\dagger c_j \sim 2 \pi \ii \int_0^\ell \dd x \left ( 1-\frac{x}{\ell} \right ) J\!\left ( x \right ) + \text{irrelevant operators},
\end{equation}
where $J(x) =\, \nord{cb}\!(x)$ is the ghost number operator.
Putting all together, we conjecture that the EH of the $c = -2$ $bc$-ghost CFT would take the form 
\begin{equation}\label{eq:conjecture}
    K_A = \int_0^\ell \dd x\, \frac{x\left (\ell - x \right )}{\ell}\, T_{00}\!\left (x\right ) + 2 \pi \ii \int_0^\ell \dd x \left ( 1-\frac{x}{\ell} \right ) J\!\left ( x \right ),
\end{equation}
which is one of the main results of this paper. 
Comparing the proposed entanglement Hamiltonian with the result for unitary CFTs in \cref{eq:CFTinterval}, the main difference is the presence of the imaginary term proportional to the ghost number $J(x)$.
Nevertheless, since this term is again the integral of a local operator, our conjecture~\eqref{eq:conjecture} retains a local structure.
Notice that, since the conformal dimension of the ghost number operator $J(x)$ is $\Delta_J = 1$, the local weight $(1-x/\ell)$ is dimensionless and it does not scale with the system size, as we observed on the lattice.

\begin{figure}[t]
    \centering
    \includegraphics[width=0.7\textwidth]{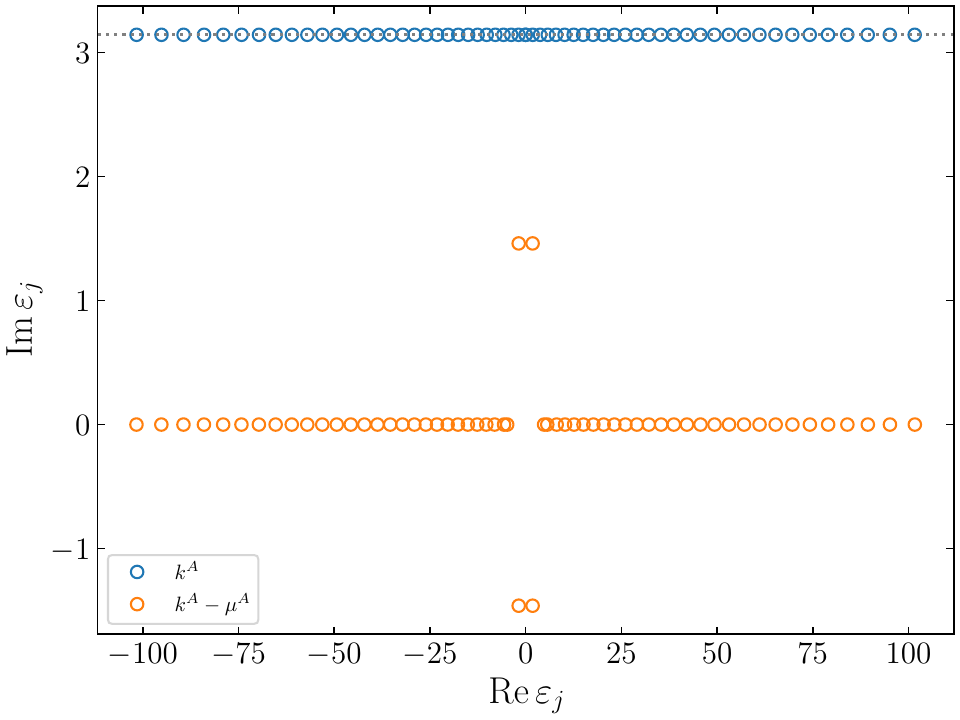}
    \caption{Spectra of the single particle EH $k^A$ (blue) and of the difference $(k^A -\mu^A)$ (orange), where $\mu^A$ is given by \cref{eq:conjecturelattice}. 
    The data are for an interval of length $\ell = 120$, in a system of size $L = 2000$, and couplings $w = 1.5$, $v = 1$ and $u = w-v = 0.5$. 
    All eigenvalues of the EH have imaginary part equal to $\pi$ (gray dotted line). Subtracting $\mu^A$ has the net effect of making almost all the eigenvalues real.}
    \label{fig:spectrumgapless}
\end{figure}

In order to understand the role played by the term $\mu^A$ in \cref{eq:conjecturelattice}, in \cref{fig:spectrumgapless} we compare the single particle entanglement spectrum, i.e., the eigenvalues of $k^A$, with the eigenvalues of the matrix $(k^A-\mu^A)$.
All the eigenvalues $\varepsilon_j$ of the single-particle entanglement Hamiltonian (blue circles)
possess an imaginary part equal to $\pi$, a feature previously identified in Ref.~\cite{Fossati:2023zyz}.
As already mentioned, this imaginary part is due to the fact that the eigenvalues $\nu_j$ of the correlation matrix all belong to $(-\infty, 0) \cup (1, +\infty)$, which, using \cref{eq:peschels}, leads to~\cite{Fossati:2023zyz}
\begin{equation}\label{eq:spectrum}
    \varepsilon_j = \log \abs{\frac{1-\nu_j}{\nu_j}} + \ii \pi\,.
\end{equation}
As discussed in \cref{sec:lrstate}, the impact of the imaginary part in \cref{eq:spectrum} on the many-body spectrum of the reduced density matrix is to impart an alternating sign to the eigenvalues of $\rho_A$ depending on the charge sector, i.e., the number of ghosts, according to~\cite{Fossati:2023zyz}
\begin{equation}
    \rho_A = \lr{-1}^{Q_A - \avg{Q_A}} \abs{\rho_A},
\end{equation}
which in turn is responsible for the negative sign of the entanglement entropy.
On the other hand, in \cref{fig:spectrumgapless} we see that the eigenvalues of $(k^A-\mu^A)$ (orange circles) are almost all real.
We can therefore argue that the novel operator $\mu^A$ in \cref{eq:conjecturelattice} (and its continuum limit~\eqref{eq:continuumconjecture} in the $bc$-ghost CFT) is the one responsible for the alternating sign of the entanglement spectrum.
Without the operator $\mu^A$,  the reduced density matrix $\rho_A$ would be positive defined and, as a consequence, the entanglement entropy would be positive too.

\section{Conclusions}\label{sec:conclusions}

In this work we have have studied the ground state entanglement Hamiltonian in the non-Hermitian SSH model, considering the left-right density matrix $\rho = \ket{L} \bra{R}$.
We studied both the topologically trivial gapped phase and the critical point.
In the gapped phase, the entanglement Hamiltonian assumes the typical triangular shape (see \cref{eq:TTISSH} and discussion) that was already observed in Ref.~\cite{Eisler:2020lyn} for unitary integrable gapped models.
Near the endpoints of the interval, the entanglement temperature grows linearly with the lattice site, according to the half-space prediction in \cref{eq:TTISSH}.
Remarkably, we observe that the same behaviour is true for the imaginary part of the entanglement Hamiltonian.
This is the first example of a lattice Bisognano-Wichmann like behaviour in a non-Hermitian model.

At the critical point, described by the $bc$-ghost CFT, we find a departure from the parabolic entanglement Hamiltonian in \cref{eq:CFTinterval} predicted by the Bisognano-Wichmann theorem for unitary CFTs.
In addition to a term proportional to the energy density with a parabolic entanglement temperature, we observe a term proportional to the number operator $c_i^\dagger c_i$ with an imaginary chemical potential interpolating between $2\pi \ii$ and $0$, cf. \cref{eq:conjecturelattice}.
This operator has a profound effect on the entanglement spectrum.
As depicted in \cref{fig:spectrumgapless}, removing the operator in \cref{eq:conjecturelattice} ensures that almost all the eigenvalues are real.
As discussed in Ref.~\cite{Fossati:2023zyz}, the imaginary part of the single particle entanglement spectrum in \cref{eq:spectrum} is responsible for the negativeness of the entanglement entropy.
If the operator in \cref{eq:conjecturelattice} were not present, the entanglement entropy would be positive.
Based on these results,   we formulate a conjecture given by \cref{eq:conjecture} for the entanglement Hamiltonian in the $bc$-ghost CFT.
Such a conjecture consists of a term analogous to the Bisognano-Wichmann EH in \cref{eq:CFTinterval} and of an imaginary chemical potential term proportional to the ghost number $J(x)$.

This paper paves the way for future investigations into the entanglement Hamiltonians of non-Hermitian models. 
Three open problems emerges very naturally. 
Firstly, in the gapped phase, it would be interesting to derive analytically the corner transfer matrix. 
As discussed in  \cref{sec:EHgapped}, this would determine the slope of the triangular entanglement temperature in \cref{fig:gappedEH} and could validate the lattice Bisognano-Wichmann behaviour. 
Secondly, it is desirable to analytically derive the entanglement Hamiltonian at the critical point, akin to the work done for free massless fermions in Ref.~\cite{Casini:2009vk}.
Thirdly, the robustness of our findings remains uncertain, such as whether the conjectured form of the entanglement Hamiltonian withstands the presence of relevant interactions.
Other unexplored research directions include understanding the EH for non-Hermitian systems that lack a real Hamiltonian spectrum.

\section*{Acknowledgements}
We are grateful to G. Di Giulio and I. Peschel for useful discussions.
The authors acknowledge support from ERC under Consolidator grant number 771536 (NEMO).


\bibliographystyle{ytphys}
\bibliography{bibliography}
\end{document}